\documentclass[a4paper,11pt]{article}
\pdfoutput=1 

\usepackage{jheppub} 

\usepackage[T1]{fontenc} 
\usepackage{graphicx}
\usepackage{latexsym}
\usepackage{amssymb,amsmath}

\renewcommand{\a}{\alpha}
\renewcommand{\b}{\beta}

\newcommand{\g}{\gamma}

\newcommand{\la}{\lambda}

\newcommand{\pa}{\partial}
\newcommand{\td}{\textrm{d}}
\newcommand{\nn}{\nonumber\\}

\def\be{\begin{equation}}
\def\ee{\end{equation}}
\def\bea{\begin{eqnarray}}
\def\eea{\end{eqnarray}}
\def\bal{\begin{align}}
\def\eal{\end{align}}
\newcommand{\bit}{\begin{itemize}}
\newcommand{\eit}{\end{itemize}}

\title{Extensions of two-field mimetic gravity}

\author[a,b,c,h,1]{Yunlong Zheng,\note{Corresponding author}}
\author[d,e,f,g]{and Haomin Rao}

\affiliation[a]{Center for Gravitation and Cosmology, College of Physical Science and Technology, Yangzhou
University, Yangzhou 225009, China}
\affiliation[b]{CAS Key Laboratory for Researches in Galaxies and Cosmology, Department of Astronomy, University of Science and Technology of China, Hefei, Anhui 230026, China}
\affiliation[c]{School of Astronomy and Space Science, University of Science and Technology of China, Hefei, Anhui 230026, China}
\affiliation[d]{School of Fundamental Physics and Mathematical Sciences, Hangzhou Institute for Advanced Study, UCAS, Hangzhou 310024, China}
\affiliation[e]{University of Chinese Academy of Sciences, 100190 Beijing, China}
\affiliation[f]{Interdisciplinary Center for Theoretical Study, University of Science and Technology of China, Hefei, Anhui 230026, China}
\affiliation[g]{Peng Huanwu Center for Fundamental Theory, Hefei, Anhui 230026, China}
\affiliation[h]{ICRANet, Piazza della Repubblica 10, I-65122 Pescara, Italy}
\emailAdd{zhyunl@ustc.edu.cn}
\emailAdd{raohaomin@ucas.ac.cn}

\abstract{
Two-field mimetic gravity was recently realized by looking at the singular  limit of the conformal transformation between the auxiliary metric and the physical metric with two scalar fields involved. In this paper, we reanalyze the singular conformal limit and find  a more general  solution for the conformal factor $A$, which greatly broadens the form of  two-field mimetic constraint and thus extends the two-field mimetic gravity. 
 We find the general setup still mimics the role of dark matter at the cosmological background level. Moreover, we extend the action by introducing extra possible term for phenomenological interests. Surprisingly, some special cases are found to be equivalent to general relativity, k-essence theory and Galileon theory. Finally, we further extend the theory by allowing the expression of mimetic constraint to be arbitrary  without imposed condition, and show that the dark matter-like behavior  is unaffected even in this extension.
}

\begin{document}
\maketitle
\flushbottom
\section{Introduction}\label{Introduction}
%
Theories of modified gravity \cite{Joyce:2014kja,Berti:2015itd,Koyama:2015vza,Cai:2015emx} have attracted considerable attention in the last few years   as  possible solutions for the dark energy, dark mater, the singularity problem \cite{Novello:2008ra,Lehners:2008vx,Cai:2014bea,Cai:2012va} and so on. Mimetic scenario was  first proposed by Chamseddine and Mukhanov \cite{Chamseddine:2013kea}  as a modification of  Einstein's general relativity where the scalar field play the role of dark matter. The idea is to express the physical metric $g_{\mu\nu}$ in the Einstein-Hilbert action in terms of an auxiliary metric $\tilde{g}_{\mu\nu}$ and a scalar field $\phi$, as follows
\begin{equation}\label{conformal}
g_{\mu\nu}=\left(\tilde{g}^{\alpha\beta}\phi_{\alpha}\phi_{\beta}\right)\tilde{g}_{\mu\nu}~,
\end{equation}
in which  $\phi_{\alpha}=\nabla_{\alpha}\phi$ is the covariant derivative of the scalar field with respect to spacetime, so that the physical metric is invariant under Weyl rescalings of $\tilde{g}_{\mu\nu}$. The above relation then leads to a constraint equation
\begin{equation}
g^{\mu\nu}\phi_{\mu}\phi_{\nu}=1~.
\end{equation}
The gravitational equations by varying the Einstein-Hilbert action, which is constructed from the physical metric $g_{\mu\nu}$, contains an extra scalar mode which can mimic the cold dark matter. 
Alternatively, one can impose the above  constraint  by employing a Lagrange multiplier \cite{Golovnev:2013jxa} in the action. The mimetic gravity written in the  Lagrange multiplier formalism takes the following form
\begin{equation}
S= \int\td^4 x \sqrt{-g} \left[\frac{1}{2} R+\frac{\lambda}{2}\left(g^{\mu\nu}\phi_{\mu}\phi_{\nu}-1 \right)+\mathcal{L} _{m}\right]~,
\end{equation}
where we have used the unit reduced Planck mass $M_p^2 = 1/(8\pi G) = 1$ and the most negative signature for the metric.

For phenomenological applications, the mimetic model  was  generalized in Ref.~\cite{Chamseddine:2014vna,Lim:2010yk} by introducing  a potential term $V (\phi)$. With appropriate choice of the potential, the generalized model  can provide inflation, bounce, dark energy, and so on.  This has stimulated  extensive cosmological and astrophysical interests \cite{Saadi:2014jfa,Mirzagholi:2014ifa,Matsumoto:2015wja,Myrzakulov:2015kda, Astashenok:2015qzw,Chamseddine:2016uyr,Nojiri:2016vhu,Babichev:2016jzg,Chamseddine:2016ktu,Shen:2017rya,Casalino:2018tcd,Dutta:2017fjw,Gorji:2020ten,Nashed:2021pkc,Farsi:2022hsy,Nashed:2022yfc,HosseiniMansoori:2020mxj,Zhang:2022bde},
 and mimetic scenario has also been applied in various modified gravity theories  
\cite{Leon:2014yua,Astashenok:2015haa,Myrzakulov:2015qaa,Rabochaya:2015haa,Arroja:2015yvd,Cognola:2016gjy,Nojiri:2016ppu,Chamseddine:2018qym,Zhong:2017uhn,Zhong:2018fdq,Xiang:2020qrc,Chamseddine:2019gjh,Liu:2017puc,Izaurieta:2020kuy}.
 The Hamiltonian analysis of different kinds of mimetic models were investigated  in Refs.~\cite{Malaeb:2014vua,Chaichian:2014qba,Takahashi:2017pje,Zheng:2018cuc,Ganz:2018mqi,Malaeb:2019rdl,Ganz:2019vre,deCesare:2020got}.  There are also some other theoretic developments  \cite{Hammer:2015pcx,Golovnev:2018icm,Langlois:2018jdg,Gorji:2018okn,Sheykhi:2020dkm}.
 See Ref.~\cite{Sebastiani:2016ras} for a review.   

Although  the original mimetic gravity is free of pathologies, it is shown that the  scalar perturbation is non-propagating due to the mimetic constraint even being offered a potential, thus can't be quantized in a usual way. Besides, this may also lead to caustic singularities. To remedy these issues, higher derivative terms of the mimetic field $(\Box\phi)^2$ are introduced \cite{Chamseddine:2014vna} to promote the scalar mode to be propagating.  The equation of motion for the scalar perturbation indeed has the wave-like form by choosing appropriate coefficient. However, the detailed analysis in the action formalism indicates that the mimetic model with higher derivatives always suffer from ghost or gradient instability at the level of linear perturbations \cite{Ijjas:2016pad,Firouzjahi:2017txv}. Then it was suggested \cite{Zheng:2017qfs,Hirano:2017zox,Gorji:2017cai} to circumvent this difficulty by considering the direct couplings of the higher derivatives of the mimetic field to the spacetime curvature.

Another interesting aspect of mimetic gravity is its close relation with non-invertible disformal  transformation. It is pointed out in Refs.~\cite{Deruelle:2014zza,Arroja:2015wpa,Domenech:2015tca} that mimetic gravity can be realized through a non-invertible disformal transformation \cite{Bekenstein:1992pj} where the number of degrees of freedom (DOFs) is no longer preserved  between the two frames \cite{Takahashi:2017zgr} and the additional DOF plays the role of dark matter. In this regard, the two-field extension of the mimetic scenario \cite{Firouzjahi:2018xob,Shen:2019nyp}
 was recently  proposed by looking at the singular  limit in the special case of conformal transformation 
 \begin{equation}\label{twoconformal}
g_{\mu\nu}=A(\phi,\psi,\tilde{X},\tilde{Y},\tilde{Z})\tilde{g}_{\mu\nu}~,
\end{equation}
where  $\tilde{X} \equiv \tilde{g}^{\mu \nu} \phi_{ \mu} \phi_{ \nu},~ \tilde{Y} \equiv \tilde{g}^{\mu \nu} \psi_{ \mu} \psi_{ \nu},~\tilde{Z} \equiv \tilde{g}^{\mu \nu} \phi_{ \mu} \psi_{ \nu}$, and the conformal factor $A$ is the function of $\phi,\psi,\tilde{X},\tilde{Y}$, and $\tilde{Z} $. 
Note that the notations used in this paper are slightly different from those of Ref.~\cite{Firouzjahi:2018xob}.
 The condition for singular limit is derived to be 
 \be\label{cd1}
	A(\phi,\psi,\tilde{X},\tilde{Y},\tilde{Z})=\tilde{X} A_{,\tilde{X}}+\tilde{Y} A_{,\tilde{Y}}+\tilde{Z} A_{,\tilde{Z}}~,
\ee
where a comma denotes a partial derivative with respect to the argument. Then the authors of \cite{Firouzjahi:2018xob} claimed that the nontrivial solution for $A$ is
 \be\label{c2}
     A=\a\tilde{X}+\b\tilde{Y}+2\g \tilde{Z}~,
 \ee
where $\a$ and $\b$ are  arbitrary functions of the scalar fields $\phi$ and $\psi$. After imposing shift symmetries on both $\phi$ and $\psi$, it was found \cite{Firouzjahi:2018xob,Shen:2019nyp} that this setup still mimics the dark matter at the cosmological background level. Recently, mimetic gravity is  extended  to the multi-field setup  in \cite{Mansoori:2021fjd}   with a curved field space manifold.

In this paper, we will reanalyze the singular limit of  conformal transformation with two scalar fields, and then extend the two-field mimetic theory. This paper is organized as follows. In section \ref{sec:constraint}, we  obtain the  more general solution for the condition on the conformal factor $A$ in singular limit, and  give several  explicit  forms  of function $A$ as examples to illustrate  new possibilities. In section \ref{sec:general}, the generalized two-field mimetic gravity is proposed and the dark matter-like fluid is confirmed. In section \ref{extensions},  we extend the theory by introducing extra possible terms and some special cases are found to be equivalent to  general relativity,  k-essence theory and Galileon theory.  In section \ref{further}, further possible extensions are proposed. Section \ref{sec:discussion} is our conclusion.

\section{Singular conformal limit and  two-field mimetic constraint}
\label{sec:constraint}
As pointed out by Refs.~\cite{Deruelle:2014zza,Arroja:2015wpa},  one can find the original single field mimetic gravity  by  performing a singular conformal transformation to the pure Einstein-Hilbert action. Similarly, two-field extension of mimetic gravity can be found  by looking at the singular limit of the conformal transformation \eqref{twoconformal} which involves two scalar fields  \cite{Firouzjahi:2018xob}.  Requiring the  eigenvalue for the Jacobian of the transformation $\frac{\pa g_{\mu\nu}}{\pa \tilde{g}_{\a\b}}$ to be zero, the condition on $A$  in the singular limit  \eqref{twoconformal} is obtained as
$A(\phi,\psi,\tilde{X},\tilde{Y},\tilde{Z})=\tilde{X} A_{,\tilde{X}}+\tilde{Y} A_{,\tilde{Y}}+\tilde{Z} A_{,\tilde{Z}}~.$
In this section, we will reanalyze the condition on the conformal factor A,  and then we derive the corresponding two-field mimetic constraint and show several  forms of $A$ to illustrate  new possibilities. 
 \subsection{Singular conformal limit}
One can write the condition \eqref{cd1} on $A$   in a  simpler way
\be \label{cd2}
  A(\phi,\psi, c \tilde{X}, c \tilde{Y}, c \tilde{Z})=c ~ A(\phi,\psi,\tilde{X},\tilde{Y},\tilde{Z})~,
\ee
which holds for any number or  scalar field $c$. We will prove the equivalence between eq.~\eqref{cd1} and eq.~\eqref{cd2} in the appendix \ref{apa}. This means  $A(\phi,\psi,\tilde{X},\tilde{Y},\tilde{Z})$ is a homogeneous function of degree one with respect
to  $\tilde{X},\tilde{Y}$ and $\tilde{Z}$.
Writing the kinetic term $\tilde{X},\tilde{Y},\tilde{Z}$ in term of  the inverse metric  and the covariant derivative of the scalar fields, i.e.,  $A(\phi,\psi,\tilde{X},\tilde{Y},\tilde{Z})=A(\phi,\psi,\phi_{\mu},\psi_{\mu}, c~ \tilde{g}^{\mu\nu})$, one immediately has
\be \label{cd3}
  A(\phi,\psi,\phi_\mu,\psi_\mu, c ~\tilde{g}^{\mu\nu})=c ~A(\phi,\psi,\phi_\mu,\psi_\mu,  \tilde{g}^{\mu\nu})~,
\ee
which means $A(\phi,\psi,\phi_\mu,\psi_\mu,  \tilde{g}^{\mu\nu})$ is a linear function with respect to the inverse metric $ \tilde{g}^{\mu\nu}$. 

The authors of \cite{Firouzjahi:2018xob} claimed that the nontrivial solution for $A$ is
 \be\label{cd4}
     A(\phi,\psi,\tilde{X},\tilde{Y},\tilde{Z})=\a(\phi,\psi) \tilde{X}+\b(\phi,\psi) \tilde{Y}+2\g(\phi,\psi) \tilde{Z}~.
 \ee
In the following subsection, several specific forms will be given to demonstrate  that the solution of eq.~(\ref{cd2}) or eq.~(\ref{cd3}) is actually far more general than eq.~(\ref{cd4}).

\subsection{Two-field mimetic constraint} 
We now move on to derive the general two-field mimetic constraint for the singular conformal transformation. According to the conformal relation between the physical metric and auxiliary metric \eqref{conformal}, one has
\be
g^{\mu\nu}=\tilde{g}^{\mu\nu}/A(\phi,\psi, \tilde{X}, \tilde{Y}, \tilde{Z})~,
\ee
and 
\be
X=\tilde{X}/A(\phi,\psi, \tilde{X}, \tilde{Y}, \tilde{Z}), ~Y=\tilde{Y}/A(\phi,\psi, \tilde{X}, \tilde{Y}, \tilde{Z}), ~Z=\tilde{Z}/A(\phi,\psi, \tilde{X}, \tilde{Y}, \tilde{Z})~.
\ee
where  $X \equiv g^{\mu \nu} \phi_{ \mu} \phi_{ \nu},~ Y \equiv g^{\mu \nu} \psi_{ \mu} \psi_{ \nu}$ and $Z \equiv g^{\mu \nu} \phi_{ \mu} \psi_{ \nu}$. Taking the value of $c$ in equation \eqref{cd2} to be $g^{\mu\nu}/\tilde{g}^{\mu\nu}$, i.e., $c=1/A(\phi,\psi, \tilde{X}, \tilde{Y}, \tilde{Z})$, one obtains
\be
	A(\phi,\psi,X,Y,Z)=A(\phi,\psi, c\tilde{X}, c\tilde{Y}, c\tilde{Z})=c~ A(\phi,\psi, \tilde{X}, \tilde{Y}, \tilde{Z})=1~.
\ee
Therefore, the general two-field mimetic constraint equation is found to be
\be \label{mc} 
	A(\phi,\psi,X,Y,Z)=1~,
\ee
which is the main result we get in this section.

\subsection{Specific forms}\label{form}
In this subsection, we will present several specific forms of the conformal factor $A$ which satisfies the condition imposed by the singular conformal limit.

\begin{itemize}
\item Form I:  $A(\phi,\psi,\tilde{X},\tilde{Y},\tilde{Z})=\a(\phi,\psi) \tilde{X}+\b(\phi,\psi)  \tilde{Y}+\g(\phi,\psi)  \tilde{Z}$ , where $\a$ and $\b$ are  arbitrary functions of the scalar fields $\phi$ and $\psi$. This means $A$ is a linear function of $\tilde{X},\tilde{Y}$ and $\tilde{Y}$ at the same time.  The associated mimetic constraint is 
\be\label{c1}
A(\phi,\psi,X,Y,Z)=\a(\phi,\psi)X+\b(\phi,\psi) Y+\g(\phi,\psi)  Z=1~.
\ee
This form has been obtained and discussed in \cite{Firouzjahi:2018xob}.
\begin{itemize}
\item Subform Ia: $\a=1,~\b=\g=0$. This case corresponds to the original single field mimetic scenario \cite{Chamseddine:2013kea}, where the conformal factor is taken to be $A(\phi,\tilde{X})=\tilde{X}$ and the associated mimetic constraint is 
\be
A(\phi,X)=X=1~.
\ee
\item Subform Ib: $\g=0$. For this case, the cross term is vanishing. Actually, one  can always remove the cross term  in \eqref{c1} through the linear transformation  of the field space. Then the conformal factor takes the form 
$ A(\phi,\psi,\tilde{X},\tilde{Y},\tilde{Z})=\a(\phi,\psi) \tilde{X}+\b(\phi,\psi) \tilde{Y}$,
and  the associated mimetic constraint is 
\be
A(\phi,\psi,X,Y,Z)=\a(\phi,\psi)X+\b(\phi,\psi) Y=1~.
\ee
If one further impose the shift symmetries on both scalar fields,  the associated mimetic constraint becomes 
\be
A(\phi,\psi,X ,Y, Z)=\a X+\b Y=1~,
\ee
 where $\a$ and $\b$ are constants. Cosmological implication of such case is discussed in detail in Refs.~\cite{Firouzjahi:2018xob,Shen:2019nyp}
\end{itemize}

\item  Form II: $A(\phi,\psi,\tilde{X},\tilde{Y},\tilde{Z})=\a(\phi,\psi) \tilde{X}^p \tilde{Y}^{q}\tilde{Z}^{{1-p-q}}$ , where $p$ and $q$ are contants. Note the conformal factor $A$ is not   a linear function of $\tilde{X},\tilde{Y}$ and $\tilde{Y}$ any more.  The associated mimetic constraint is 
\be\label{c2}
A(\phi,\psi,X,Y,Z)=\a(\phi,\psi)X^p Y^q Z^{1-p-q}=1~.
\ee
This form hasn't been discussed yet in the literature, and we hope to come back to study its cosmology soon.

\begin{itemize}
\item Subform IIa: $p+q=1$ and $\a(\phi,\psi)=1$. The conformal factor takes the form $A(\phi,\tilde{X})= \tilde{X}^p \tilde{Y}^{1-p}$ and the associated mimetic constraint is $A(\phi,X)=X^p Y^{1-p}=1$. One interesting possibility is to choose  $p=1/(m+1)$ and $q=m/(m+1)$ where $m$ is an positive integer. The mimetic constraint then reads $X^{1/(m+1)} Y^{m/(1+m)}=1$, which yields
\be
X Y^{m}=1~.
\ee
For instance, choosing $m=1$ leads to the mimetic constraint $X Y=1$. 
Another interesting choose is to take $p=n+1$ and $q=-n$ ($n$ is also an positive integer), which gives the conformal factor $A(\phi,\psi,\tilde{X},\tilde{Y},\tilde{Z})=\tilde{X}^{n+1} /\tilde{Y}^{n}$ and the mimetic constraint  
\be
X^{n+1} /Y^{n}=1~.
\ee
Fox instance, choosing $n=1$ gives the mimetic constraint  equation  $X^2/  Y=1$. 
\item Subform IIb:   $p+q\neq1$. For instance, choosing $p=q=1$  the mimetic constraint  equation becomes 
\be
X Y/Z=1~.
\ee 
\end{itemize}

\item  Form III: $A(\phi,\psi,\tilde{X},\tilde{Y},\tilde{Z})=\Sigma_{i} ~\a_{i}(\phi,\psi) A_{i}(\phi,\psi,\tilde{X},\tilde{Y},\tilde{Z})$, where $\a_{i}$ are arbitrary functions of $\phi$ and $\psi$, and $A_{i}$  satisfy the condition imposed by singular conformal limit. This means that any linear combination of the conformal factors  satisfying the singular limit condition \eqref{cd1} still safisfies  \eqref{cd1}, which is a natural result of the fact that equation  \eqref{cd1} is a linear equation of $A$. To show  new possibilities  of this form,  here we give a concrete example where the constraint equation is taken to be
\be
A(\phi,\psi,X,Y,Z)=X+Y+\a X^2/Y=1~.
\ee
\end{itemize}

Actually, one can prove all the solution of $A$ in the singular conformal limit can be written in the general form: 
\be\label{form3}
A(\phi,\psi,\tilde{X},\tilde{Y},\tilde{Z})=\tilde{X} \theta(\phi,\psi,\tilde{Y}/\tilde{X},\tilde{Z}/\tilde{X})~,
\ee
 where $\theta$ is an arbitrary function of $\phi,\psi,Y/X$, and $Z/X$. 
For this general form of $A$,  the mimetic constraint  equation reads 
\be\label{c3}
X \theta(\phi,\psi,Y/X,Z/X)=1~.
\ee 
For example, one can rewrite  form I and  form II as 
\be
A(\phi,\psi,\tilde{X},\tilde{Y},\tilde{Z})=\tilde{X}\left(\a(\phi,\psi) +\b(\phi,\psi) \frac{\tilde{Y}}{\tilde{X}}+\g(\phi,\psi)  \frac{\tilde{Z}}{\tilde{X}}\right)~, 
\ee
and 
\be
A(\phi,\psi,\tilde{X},\tilde{Y},\tilde{Z})= \tilde{X}\left[\a(\phi,\psi)  \left(\frac{\tilde{Y}}{\tilde{X}}\right)^q  \left(\frac{\tilde{Z}}{\tilde{X}}\right)^{1-p-q}\right]~,
\ee
which obviously satisfy the form \eqref{form3}.  To illustrate how  general  the form of two-field mimetic constraint  can be, let us show three more explicit cases: 
\be\label{c32}
X e^{-Y/X}=1~, ~\frac{X}{ \sin{(Y/X)}}=1~, ~\frac{X}{ \sin{(Y/X)} e^{-\phi}}=1~.
\ee 

\section{General  two-field mimetic gravity}\label{sec:general}
Similar to the original mimetic scenario, the action for two-field mimetic gravity can be written as
\be\label{action0}
 S[\tilde{g}_{\mu\nu},\phi,\psi]=\int\td^4x\sqrt{-g(\tilde{g}_{\mu\nu},\phi,\psi)}\frac{R(g_{\mu\nu}(\tilde{g}_{\mu\nu},\phi,\psi))}{2}~,
 \ee
where the  relation between the physical metric  and the auxiliary metric is \eqref{twoconformal} and the physical metric is invariant under the Weyl rescalings of the auxiliary metric.  The conformal factor $A$ in the singular limit satisfies the condition \eqref{cd1}, or equivalently, \eqref{cd2}, and the associated mimetic constraint is given by \eqref{mc}. It is convenient to rewrite the action in the Lagrange multiplier formalism and we have
\be\label{action1}
 S[g,\phi,\psi,\la]=\int\td^4x\sqrt{-g}\left[\frac{R}{2}+\frac{\la}{2} \bigg(A(\phi,\psi,X,Y,Z)-1\bigg)\right]~,
\ee
where $\la$ is  a Lagrange multiplier which enforces the mimetic constraint. Recall that  function $A(\phi,\psi,X,Y,Z)$ satisfies the condition
 \be\label{cd}
	A=X A_{,X}+YA_{,Y}+ZA_{,Z}~,
\ee
or
\be
  A(\phi,\psi,c X,c Y, c Z)=c ~ A(\phi,\psi,X,Y,Z)~,
\ee
in which we have replaced the auxiliary metric in \eqref{cd1} and \eqref{cd2} by physical metric. Several allowed forms of $A$ have been shown in the previous section. 

\subsection{Equations of motion}
This subsection is devoted to the equations of motion for  general  two-field mimetic gravity \eqref{action1}. Varying the action with respect to the metric gives the gravitational equations
\be
G_{\mu\nu}=-T_{\mu\nu}~,
\ee
where the effective energy-momentum tensor for the double mimetic fields is
\be\label{tuv}
T_{\mu\nu}=\la\left(A_{,X}\phi_\mu\phi_\nu+A_{,Y}\psi_\mu\psi_\nu+\frac{A_{,Z}}{2}(\phi_\mu\psi_\nu+\phi_\nu\psi_\mu)\right)~.
\ee
The divergence of the gravitational equations leads to
\be\label{divt}
\nabla_{\mu}T^{\mu\nu}=0~,
\ee
which represents the  conservation of energy and momentum. Varying the action with respect to the scalar fields  yields
 \be\label{eomscalar}
\nabla_{\mu}\left[\la (A_{,X}\phi^\mu+\frac{A_{,Z}}{2}\psi^\mu)\right]-\frac{\la}{2} A_{,\phi}=0~,
\nabla_{\mu}\left[\la (A_{,Y}\psi^\mu+\frac{A_{,Z}}{2}\phi^\mu)\right]-\frac{\la}{2} A_{,\psi}=0~.
 \ee
 In addition, the mimetic constraint equation $ A(\phi,\psi,X,Y,Z)=1$  is imposed by the Lagrangian multiplier.
 
 \subsection{Cosmological implications} 
 In this subsection, we consider the cosmological implications of  general two-field mimetic gravity  \eqref{action1} at the background level. For a spatially flat Friedmann-Robertson-Walker (FRW) background, the metric is given by
 \be
 ds^2=dt^2-a(t)^2 \delta_{ij}dx^i dx^j~,
 \ee
 where $a(t)$ is the scale factor and t is the cosmic time.  The energy density is
 \be
 \rho=\la(A_{,X}\dot{\phi}^2+A_{,Y}\dot{\psi}^2+A_{,Z}\dot{\phi}\dot{\psi})~,
 \ee
 and pressure is 
  \be
 \quad p=0~,
 \ee
 by computing the diagonal components of the energy-momentum tensor  \eqref{tuv}. Using the condition on $A$ 
 \be
 A=XA_{,X}+YA_{,Y}+ZA_{,Z}=(A_{,X}\dot{\phi}^2+A_{,Y}\dot{\psi}^2+A_{,Z}\dot{\phi}\dot{\psi})~,
 \ee
 together with the mimetic constraint equation
 \be
 A=1~,
 \ee
 we find $\rho=\la$ and $p=0$. The conservation of energy yields
 \be
 \dot{\rho}+3H(\rho+p)=0~,
 \ee
 in which $H=\dot{a(t)}/a(t)$ is the Hubble expansion rate. As the pressure is zero, one get 
 \be\label{rho}
 \rho=\la=c/ a^{3}~,
 \ee
where $c$ is  constant of integration. This result can also be derived from the scalar equations \eqref{eomscalar} at the cosmological background. 

In Ref.~\cite{Firouzjahi:2018xob}, the authors found that  the two-field mimetic setup after imposing the shift symmetries on both $\phi$ and $\psi$ mimics the dark matter at the cosmological background level. Here we have greatly generalized their results. Although the functional form of $A(\phi,\psi,X,Y,Z)$  in our more general setup \eqref{action1}  could be very  diverse as shown in the subsection \ref{form}, the model still describes a  dark matter-like fluid   at the cosmological background. This  means the dark matter nature is independent of the explicit form of $A$, and it is not necessary to assume shift symmetry to obtain the dark matter-like behavior. Note that  there is an exceptional case:  $\la=0$, where the mimetic scalars field don't contribute at all and the setup reduces to general relativity. We will encounter and study such case in the next section.  
 The two-field mimetic action \eqref{action1}  at the perturbation level  will be studied in detail in future work.

\section{Extension with introduced extra  term and its relations with other modified  theories of gravity}\label{extensions}
For the original mimetic scenario,  the most general term (without higher derivative) one can introduce to the scalar field is a potential term $V(\phi)$ as the kinetic term satisfies the mimetic constraint $X=1$ and thus can be eliminated. For the two-field mimetic gravity model \eqref{action1}, one can also introduce extra possible term for  phenomenology so that the extended  action takes the following form
\be\label{action3}
 S=\int\td^4x\sqrt{-g}\left[\frac{R}{2}+\frac{\la}{2} \bigg(A(\phi,\psi,X,Y,Z)-1\bigg)+Q(\phi,\psi,X,Y,Z)\right]~,
\ee
where $Q$ is an arbitrary function of $\phi,\psi,X,Y$ and $Z$.
We should emphasize that  the above extension, similar to the case of original mimetic scenario with extra potential term, in general don't mimic dark matter any more. This is because  the pressure $p=Q(\phi,\psi,X,Y,Z)$ no longer vanishes at the cosmological background level.

Based on whether one of the scalar fields can be  eliminated by the mimetic constraint, we can  divide  the above action into two categories. For our interest in this paper, we  focus more on  the cases where one of the scalar fields can  be removed  and show the possible  relations with other modified theories of gravity.

\subsection{Case where  $\psi$ can be eliminated}\label{elimianted}
Without loss of generality, we consider the case $\psi$ can be eliminated. If we choose $A$ in the action \eqref{action3} to be $X/J(\phi,\psi)$,which obviously satisfy the condition \eqref{cd},  the mimetic constraint becomes
\be\label{case1}
  A=X/J(\phi,\psi)=1~,
\ee
or equivalently,
\be\label{case12}
 \psi=f(\phi,X),
\ee
where $f$ is the inverse function of $J$.  Note that we have assumed $J(\phi,\psi)$ explicitly depending on $\psi$, i.e., $J_\psi $ is not vanishing, to ensure the existence of the inverse function. Thus the action becomes
\be \label{action32}
 S=\int\td^4x\sqrt{-g}\left[\frac{R}{2}+\frac{\tilde{\la}}{2}\bigg (\psi-f(\phi,X)\bigg)+Q(\phi,\psi,X,Y,Z)\right]~.
\ee
It should be mentioned that  the Lagrange multiplier are  redefined as  the mimetic constraint equation has been rewritten.  Variation of the above action with respect to $\tilde{\la}$ yields the mimetic constraint \eqref{case12}. Then substituting the constraint into the action, one can eliminate the scalar field $\psi$.
Therefore,  this setup  reduces to the scalar-tensor theory with only one scalar filed and no mimetic constraint at all. However, higher derivative of the remaining scalar field may comes out. We will consider some particular cases below  according to the explicit  form of the term $Q(\phi,\psi,X,Y,Z)$.

\subsubsection{Subcase without introducing the term $Q$}
Considering the subcase without introducing the extra  term $Q$, i.e.,  one has
\be \label{action33}
 S=\int\td^4x\sqrt{-g}\left[\frac{R}{2}+\frac{\la}{2} \left(\frac{X}{J(\phi,\psi)}-1\right)\right]~,
\ee
or equivalently
\be \label{action34}
 S=\int\td^4x\sqrt{-g}\left[\frac{R}{2}+\frac{\tilde{\la}}{2}\bigg (\psi-f(\phi,X)\bigg)\right]~.
\ee
The mimetic constraint $\psi-f(\phi,X)=0$ is enforced by the Lagrange multiplier. Then substituting the constraint equation into the action, the action reduces to the Einstein-Hilbert action
\be \label{action35}
 S=\int\td^4x\sqrt{-g} ~\frac{R}{2}~,
\ee
and one recovers Einstein's theory of general relativity. Therefore, we find the Weyl scaling invariant two-field mimetic theory \eqref{action0} with 
\be
g_{\mu\nu}=\tilde{X}/J(\phi,\psi)\tilde{g}_{\mu\nu}
\ee
 is equivalent to general relativity and the number of degrees of freedom in this setup is $2$.
 
 As we have concluded in section  \ref{sec:general},  the general two-field mimetic model \eqref{action1} are expected to mimic the dark matter. The special case \eqref{action33} obviously belongs to the action \eqref{action1}, but no dark matter appears. This is because the setup   \eqref{action33} is actually the exceptional case $\la=0$ we have mentioned earlier. One can verify this by varying the action    with respect to $\psi$.

\subsubsection{Subcase with $Q=Q(\phi,\psi,X)$}
Here we consider the subcase with $Q=Q(\phi,\psi,X)$ where $Q$ is explicitly dependent on $\phi,\psi$ and $X$. One has
\be \label{action36}
 S=\int\td^4x\sqrt{-g}\left[\frac{R}{2}+\frac{\tilde{\la}}{2} \bigg(\psi-f(\phi,X)\bigg)+Q(\phi,\psi,X)\right].
\ee
By using the two-field mimetic constraint one can reduce the action to 
\be 
 S=\int\td^4x\sqrt{-g}\left[\frac{R}{2}+K(\phi, X)\right]~,
\ee
where $K(\phi, X)\equiv Q(\phi,f(\phi,X),X)$. Thus  this subcase of two-field mimetic  gravity is equivalent to K-essence theory \cite{Armendariz-Picon:2000ulo}, which exists 3 degrees of freedom and has various phenomenological implications in astrophysics and cosmology. 

\subsubsection{subcase with $Q=G_2(\phi,X)-Z$}
We consider the case with $Q=G_2(\phi,X)-Z$ where $G_2$ is an arbitrary function of $\phi$ and $X$. The action becomes
\be  \label{action37}
 S=\int\td^4x\sqrt{-g}\left[\frac{R}{2}+\frac{\tilde{\la}}{2} \bigg(\psi-f(\phi,X)\bigg)+G_2(\phi,X)-\phi^\mu\psi_\mu\right].
\ee
By using the two-field mimetic constraint,  the action reduces to 
\be  
 S=\int\td^4x\sqrt{-g}\left[\frac{R}{2}+G_2(\phi,X)+G_3(\phi, X)\Box\phi\right]~,
\ee
where $G_3(\phi, X)=f(\phi,X)$. Surprisingly,  this subcase corresponds to a type of the Galileon theory \cite{Deffayet:2011gz}, which also exist 3 degrees of freedom and has numerous phenomenological implications.

\subsubsection{subcase with $Q=Q(Y)$}
We consider the case when the function $Q$  only depends on  $Y=\psi^\mu\psi_\mu$. After employing the 
mimetic constraint, one has $Y=\nabla_{\mu}f(\phi,X) \nabla^{\mu}f(\phi,X)$.  Thus, the action 
\be  \label{action38}
 S=\int\td^4x\sqrt{-g}\left[\frac{R}{2}+\frac{\tilde{\la}}{2} \bigg(\psi-f(\phi,X)\bigg)+ Q(Y)\right].
\ee
reduces to 
\be
 S=\int\td^4x\sqrt{-g}\left[\frac{R}{2}+Q(\nabla_{\mu}f\nabla^{\mu}f) \right]~,
\ee
which includes  higher derivative of the scalar field $\phi$.  This case has 4 degrees of freedom and one scalar degree of freedom  suffers from Ostrogradski instability, and thus not viable for phenomenology.

\subsection{Case where neither of the fields can be eliminated}
Usually one can not eliminate any scalar field by the mimetic constraint \eqref{mc}, so the  form of the action \eqref{action3} can be very diverse and has many possibilities. Here we discuss one special case which assumes the shift symmetry for both scalar field $\phi$ abd $\psi$, and the action takes the form
\be
 S=\int\td^4x\sqrt{-g}\left[\frac{R}{2}+\frac{\la}{2} \bigg(A(X,Y,Z)-1\bigg)+Q(X,Y,Z)\right]~,
\ee
where only kinetic terms are relevant.  If requiring $A$ to be a linear function of $X$,$Y$ and $Z$, the action reduces to
\be
 S=\int\td^4x\sqrt{-g}\left[\frac{R}{2}+\frac{\la}{2} \big(\a X+\b Y-1\big)+Q(X,Y,Z)\right]~.
\ee
The case with $Q=0$ has been investigated in \cite{Firouzjahi:2018xob,Shen:2019nyp} and it was found the adiabatic mode is frozen in the perturbation level. Then some interesting and important question can be asked. For instance,  whether the adiabatic mode is still frozen or become propagating after adding the extra term $Q$? If  the adiabatic mode become propagating, does it suffer from instabilities or not?  Can  inflation be  realized in such case?  We leave these questions for  future work.

\subsection{Degrees of freedom and Ostrogradski ghost}
It is important to study  the degrees of freedom for the extended two-field mimetic action \eqref{action3}.  Our main purpose  of this subsection is to illustrate that the  number of degrees of freedom depends on the form of both $A$ and $Q$. The detailed Hamiltonian analysis  is beyond the scope of this paper and is left for future investigation. 

We first consider the case with $Q=0$. If choose $A=X$, the original one-field mimetic gravity is recovered and the number of degrees of freedom is 3. If one choose $A=X/J(\phi,\psi)$ where one of the scalar fields can be eliminated by the mimetic constraint equation, the number of degrees of freedom will be 2. Or  if we take $A=\a X+\b Y$ \cite{Firouzjahi:2018xob,Shen:2019nyp}
,   the number of degrees of freedom will become 4. This implies that the degrees of freedom for the extended two-field mimetic action \eqref{action3} depends on the form of $A$. 
The above discussion in subsection \ref{elimianted} on the particular case \eqref{action32}, where one of the scalar fields can be eliminated by the mimetic constraint equation, shows that the theory possess 2, 3 and 4 degrees of freedom for the  subcases \eqref{action33},  \eqref{action36} and \eqref{action38} respectively.  Therefore, the  number of degrees of freedom also depends on the form of  $Q$.

Now we discuss the possibility of  Ostrogradski ghost. If the theory has 4 degrees of freedom, of which two are scalar modes, then we should be careful as one of the scalar modes may suffer from the Ostrogradski ghost. For instance, there is one Ostrogradski ghost for the case \eqref{action38} and no Ostrogradski ghost for the case where $A=\a X+\b Y$ and $Q=0$, although these two cases both have 4 degrees of freedom.

\section{Further extension}\label{further}
In  the previous sections, we have greatly extend two-field mimetic gravity by  obtaining a more general solution for the conformal factor $A$ in the singular limit. Moreover, possible extra term are proposed for phenomenological interest in section \ref{extensions}. This section is devoted to explore further extension. 

Note that  as the two-field mimetic action in the Lagrange multiplier formalism \eqref{action1} originates from the Weyl invariant gravity \eqref{action0}, the allowed expression of mimetic constraint is restricted by the singular limit condition.  But if  we directly start with the setup \eqref{action1}  instead of the Weyl invariant gravity, there will be  no imposed condition on the explicit form of the mimetic constraint.Thus, the  two-field mimetic model \eqref{action1} can be further generalized  as
\be\label{action4}
 S=\int\td^4x\sqrt{-g}\left[\frac{R}{2}+\frac{\la}{2} ~M(\phi,\psi,X,Y,Z) \right]~,
\ee
where $M$ is an arbitrary function of $\phi,\psi,X,Y$ and $Z$.  The corresponding mimetic constraint is 
\be
M(\phi,\psi,X,Y,Z)=0~,
\ee
 and the  Einstein equations of motion is  $G_{\mu\nu}=-T_{\mu\nu}$ in which the effective energy-momentum tensor  is given by
\be
T_{\mu\nu}=\la\left(M_{,X}\phi_\mu\phi_\nu+M_{,Y}\psi_\mu\psi_\nu+\frac{M_{,Z}}{2}(\phi_\mu\psi_\nu+\phi_\nu\psi_\mu)\right)~.
\ee
For the cosmological background, the energy density is
\be
\rho=\la\left(X M_{,X}+Y M_{,Y}+Z M_{,Z}\right)~,
\ee
and 
the pressure is $p=0$. Then the energy conservation law in an expanding universe  implies $\rho \propto a^{-3}$, which means  the  model \eqref{action4} with generalized mimetic constraint still mimics the role of dark matter at the cosmological background level.

\section{Conclusion and discussions} \label{sec:discussion}
The original mimetic scenario can be achieved by considering the conformal transformation in the singular limit. Similarly, two-field mimetic gravity has been proposed in \cite{Firouzjahi:2018xob} by looking at the singular limit of the conformal transformation between the auxiliary and the physical metrics with two scalar fields.  

In this paper, we reanalyze the singular conformal limit and obtain  a more general  solution for the conformal factor $A$. The associated constraint equation is derived to be $A(\phi,\psi,X,Y,Z)=1$, which extends  the two-field mimetic constraint and  generalize the two-field mimetic gravity  in the literature.  To illustrate  new possibilities, several allowed forms of $A$ are given and discussed. Then we consider the cosmological implications and find  our extended setup still behaves as dark matter at the background level, which means the dark matter nature is independent of the explicit form of $A$. Moreover,  extra possible term $Q$ are added to the action for phenomenological applications. Surprisingly,  general relativity, k-essence theory and Galileon theory are recovered in some particular cases where one of the scalar fields can be eliminated by the mimetic constraint equation. Finally, we show that the dark matter nature is  still, again, unchanged even in the further extension of  the two-field mimetic gravity which imposes no condition on the expression of mimetic constraint. 
We leave the detailed investigation of the linear perturbation theory  for future work. 

We have shown two different ways to extend two-field mimetic gravity  \eqref{action1},  namely \eqref{action3} and  \eqref{action4}. It is interesting to search for more possible extensions. For instance, one may  combine two extensions together, then the action becomes
 \be\label{equation42}
 S=\int\td^4x\sqrt{-g}\left[\frac{R}{2}+\frac{\la}{2} ~M(\phi,\psi,X,Y,Z)+Q(\phi,\psi,X,Y,Z)\right]~.
\ee
We may also introduce higher derivative terms of the scalar fields to the action, such as
 \be\label{equation42}
 S=\int\td^4x\sqrt{-g}\left[\frac{R}{2}+\frac{\la}{2} ~\bigg(A(\phi,\psi,X,Y,Z)-1\bigg)+Q(\phi,\psi,X,Y,Z,\Box\phi,\Box\psi)\right]~.
\ee
The  theoretic properties and cosmological applications of these extended models deserve detailed investigations in the future.

\acknowledgments
We would like to thank Mingzhe Li,  Yi-Fu Cai, Xian Gao and Shi Pi for useful discussions and comments.  This work is supported in part by  NSFC under Grant No. 11847239,  12075231 and 11961131007, by National Key R\&D Program of China (2021YFC2203100), and by the Fundamental Research Funds for the Central Universities. YZ is grateful for the hospitality of ICRANet during his visit when part of this work was carried out.

\appendix
\section{The Condition on A in the singular limit} \label{apa}
In this appendix, our purpose is to prove that eq.~\eqref{cd1} and eq.~\eqref{cd2} is the equivalent description for the condition on A in the singular limit. First, one can define 
\be
W(c)=A(\phi,\psi,c \tilde{X},c \tilde{Y},c \tilde{Z})~.
\ee
Take $c=1$ and one has 
\be\label{w}
W(1)=A(\phi,\psi, \tilde{X}, \tilde{Y}, \tilde{Z})~.
\ee 
The derivative of the function $W$ with respect to $c$ gives
\be\label{dw}
W'(c)=\tilde{X} \frac{\partial A(\phi,\psi,c \tilde{X},c \tilde{Y},c \tilde{Z})}{\partial (c \tilde{X})} +\tilde{Y} \frac{\partial A(\phi,\psi,c \tilde{X},c \tilde{Y},c \tilde{Z})}{\partial (c \tilde{Y})}  +\tilde{Z} \frac{\partial A(\phi,\psi,c \tilde{X},c \tilde{Y}, c \tilde{Z})}{\partial (c \tilde{Z})}  ~.
\ee

Then we will deduce  equation \eqref{cd2} from equation \eqref{cd1}.  Replacing $\tilde{X},\tilde{Y}$ and $\tilde{Z}$ by $c \tilde{X}, c \tilde{Y}$ and $c \tilde{Z}$ respectively, equation \eqref{cd1} becomes
\begin{align}
A(\phi,\psi,c \tilde{X},c \tilde{Y},c \tilde{Z})=& c~\tilde{X} \frac{\partial A(\phi,\psi,c \tilde{X},c \tilde{Y},c \tilde{Z})}{\partial (c \tilde{X})} +c~ \tilde{Y} \frac{\partial A(\phi,\psi,c \tilde{X},c \tilde{Y},c \tilde{Z})}{\partial (c \tilde{Y})}  \nn
&+c~\tilde{Z} \frac{\partial A(\phi,\psi,c \tilde{X},c \tilde{Y}, c \tilde{Z})}{\partial (c \tilde{Z})} ~.
\end{align}
Employing the expression of the function $W$  \eqref{w} and the its derivative \eqref{dw}, the above equation yields
\be
W(c)=c~ W'(c)~,
\ee
which means $W(c)$ is a linear function of $c$, i.e., 
\be
W(c)=c ~W(1)~.
\ee
 Using the definition of the function $W$ again, we find 
$A(\phi,\psi,c \tilde{X},c \tilde{Y},c \tilde{Z})=c~ A(\phi,\psi, \tilde{X}, \tilde{Y}, \tilde{Z})$, which is exactly equation \eqref{cd2}.

Now we  show how equation \eqref{cd1} can also be derived from equation \eqref{cd2}. Note that  \eqref{cd2} can be rewritten with function $W$, i.e., $W(c)=c ~W(1)$. Taking the derivative of both sides with respect to $c$ gives 
\be
W'(c)=W(1)~.
\ee
Using  equation \eqref{dw} and set $c$ to $1$, we obtain equation \eqref{cd1}. Thus, the equivalence of eq.~\eqref{cd1} and eq.~\eqref{cd2} is proved.


\end{document}